\documentclass{pasj00}

\begin{document}
\SetRunningHead{T. Kato et al.}{V803 Cen: Helium Dwarf Nova}

\Received{}
\Accepted{}

\title{V803 Centauri: Helium Dwarf Nova Mimicking a WZ Sge-Type Superoutburst}

\author{Taichi \textsc{Kato}}
\affil{Department of Astronomy, Kyoto University,
       Sakyo-ku, Kyoto 606-8502}
\email{tkato@kusastro.kyoto-u.ac.jp}

\author{Rod \textsc{Stubbings}}
\affil{19 Greenland Drive, Drouin 3818, Victoria, Australia}
\email{stubbo@sympac.com.au}

\author{Berto \textsc{Monard}}
\affil{Bronberg Observatory, PO Box 11426, Tiegerpoort 0056, South Africa}
\email{LAGMonar@csir.co.za}

\author{Neil D. \textsc{Butterworth}}
\affil{24 Payne Street, Mt. Louisa, Townsville, Queensland 4814, Australia}
\email{neilbutt@bigpond.com.au}

\author{Greg \textsc{Bolt}}
\affil{295 Camberwarra Drive, Craigie, Western Australia 6025, Australia}
\email{gbolt@iinet.net.au}

\email{\rm{and}}

\author{Tom \textsc{Richards}}
\affil{Woodridge Observatory, 8 Diosma Rd, Eltham, Vic 3095, Australia}
\email{tom@woodridgeobsy.org}


\maketitle

\begin{abstract}
   We observed the long-term behavior of the helium dwarf nova V803 Cen,
and clarified the existence of at least two distinct states
(a state with 77-d supercycles and a standstill-like state) that
interchangeably appeared with a time-scale of 1--2 yr.
We also conducted a time-resolved CCD photometry campaign
during the bright outburst in 2003 June.  The overall appearance of the
outburst closely resemble that of the late stage of the 2001 outburst
of WZ Sge, consisting of the initial peak stage (superoutburst plateau),
the dip, and the oscillating (rebrightening) states.
During the initial peak stage, we detected a large-amplitude
superhump-type variation with a period of 0.018686(4) d = 1614.5(4) s,
and during the oscillation stage, we detected variations with a period
of 0.018728(2) d = 1618.1(2) s.  We consider that the former period
better represents the superhump period of this system, and the latter
periodicity may be better interpreted as arising from late superhumps.
The overall picture of the V803 Cen outburst resembles
that of a WZ Sge-type outburst, but apparently with a higher
mass-transfer rate than in hydrogen-rich WZ Sge-type stars.
We suggest that this behavior may be either the result of difficulty
in maintaining the hot state in a helium disk, or the effect of
an extremely low tidal torque resulting from the extreme mass ratio.
\end{abstract}

\section{Introduction}

   Cataclysmic variables (CVs) are interacting binaries which contain
a white dwarf as the mass-accretor.  Among CVs, there are a small group
containing a helium white dwarf as the mass donor.  These systems are
either called AM CVn stars (\cite{war95amcvn}; \cite{sol95amcvnreview})
or interacting binary white dwarfs (IBWDs) [see \citet{nel01amcvnspot}
for the recent progress in confirming the binary nature].
Only a limited number of IBWDs are known at present:
AM CVn (\cite{fau72amcvn}; \cite{war72amcvn}; \cite{rob75amcvn};
\cite{sol84amcvn}; \cite{sol98amcvnWET}; \cite{pat92amcvn}),
CR Boo (\cite{nat84crbooiauc}; \cite{woo87crboo}; \cite{pat97crboo};
\cite{pro97crboo}),
V803 Cen (\cite{odo87v803cen}; \cite{odo89v803cen}; \cite{odo90v803cenWET}),
GP Com (\cite{bur71gpcom}; \cite{war72gpcom}; \cite{nat81gpcom}),
HP Lib (\cite{odo94hplib}; \cite{pat02hplib}),
CP Eri (\cite{abb92alcomcperi}; \cite{gro01cperi}),
KL Dra (\cite{jha98kldraiauc6983}; \cite{woo02kldra}),
V396 Hya = CE$-$315 (\cite{rui01ce315}; \cite{wou01v359cenxzeriyytel}),
ES Cet = KUV 01584$-$0939 (\cite{war02kuv01584}),
and SN 2003aw (\cite{wou03sn2003awiauc8085}).

   The evolution of IBWDs have been widely discussed as decendants from
a narrow channel of close binary evolution (cf. \cite{web84IBWDrcrbSNIa};
\cite{tut96IBWD}; \cite{nel01amcvnpopulationsynthesis};
\cite{yun02LMXBamcvnevolution}; \cite{pod03amcvn}),
particularly in relation to possible precursors of type-Ia supernovae
and unusual objects such as R CrB-type stars,
and in relation to recently discovered short-period unusual
X-ray sources
(\cite{cro98v407vul}; \cite{ram00v407vul}; \cite{kin02m31SSS};
\cite{mar02v407vul}; \cite{wu02v407vul}; \cite{ram02j0806};
\cite{isr02j0806}; \cite{ram02v407vul}; \cite{nor02v407vulj0806}).
A recent summary in relation to magnetic CVs can be found in
\citet{war03MCV}.

   The recent discovery of an extremely short-period, hydrogen-rich
binary (EI Psc = 1RXS J232953.9+062814) with a massive secondary
\citep{tho02j2329} seems to first observationally confirm the existence
of the predicted evolutionary path forming IBWDs \citep{uem02j2329letter}.
Together with the discovery of a shortest period CV-type binary
\citep{war02kuv01584}, the field of IBWDs provides one of the
hottest topics in modern stellar astrophysics.  These objects are
also considered to be excellent candidates for next generation
experiments of directly detecting gravitational wave radiation
(\cite{hil00amcvnGR}; \cite{str02v407vuldecay}).

   V803 Cen is one of the first identified IBWDs (\cite{odo87v803cen};
\cite{odo89v803cen}), which had been originally considered to be
an R CrB-type hydrogen-deficient variable star.  It took, however,
rather long time before the dwarf nova-type behavior of this object
was established.  \citet{war95amcvn} originally considered that this
object may be a helium analog of a VY Scl-type CV
with occasional fadings (cf. \cite{war74vysclzcam};
\cite{gar88vyscldqher}; \cite{gre98vyscl}; \cite{lea99vyscl};
\cite{kat02kraur}).
The ``low state" description \citep{odo90v803cenWET} of the object's
state exactly follows the notation in VY Scl-type CVs.  It was only
after the ``northern counterpart" CR Boo was found to more or less display
common characteristics with the hydrogen-rich dwarf novae
\citep{pat97crboo}.

   The assertion of CR Boo as being a ``dwarf nova" in \citet{pat97crboo}
was, however, later found to be an incomplete description of the behavior
what this object displays.  \citet{kat00crboo} revealed that the CR Boo is
actually a helium counterpart of an SU UMa-type dwarf nova with a very
short (46.3 d) supercycle, corresponding to ER UMa stars in hydrogen-rich
CVs (\cite{kat95eruma}; \cite{rob95eruma}; \cite{mis95PGCV};
\cite{nog95rzlmi}).  The variations \citet{pat97crboo} referred to as
dwarf nova-type outbursts were actually more unusual short-term
oscillations seen in its ``standstill-like" states
(\cite{kat00crboo}; \cite{kat01v803cen}).
The true dwarf nova-type character of CR Boo manifests itself more in the
46.3-d supercycle, which agrees quite well with the theoretical
interpretation of the disk-instabilities in a helium accretion disk
\citep{tsu97amcvn}.  \citet{kat01v425cas} suggested the possibility that
the short-term oscillations reported in \citet{pat97crboo} may
reflect more or less stablized thermal disk-instability.

   Following this progress in CR Boo, the dwarf nova-type nature of
V803 Cen was finally elucidated by the following two works:
\citet{pat00v803cen} primarily reported on the presence of superhumps
and oscillating states, and \citet{kat00v803cen} reported on the detection
of the supercycle almost identical with that of CR Boo.  The oscillating
state in \citet{pat00v803cen} is likely a ``standstill"
\citep{kat01v803cen}, which is characteristics of Z Cam-type dwarf novae
(see e.g. \cite{hel01book} Sect. 5.4; see also \cite{war74vysclzcam};
\cite{mey83zcam}; \cite{opp98zcam}).  Since there is no known hydrogen-rich
object which shows both superoutbursts and standstills, this feature of
CR Boo and V803 Cen is quite challenging to dwarf nova theories.
Furthermore, \citet{kat01crboo} reported a sudden dramatic change in the
supercycle in CR Boo, which is also unexpected from the analogy with
hydrogen-rich CVs.  Both aspects will be discussed in later sections.

\section{Observation}

   The long-term visual observations were mostly undertaken by
Rod Stubbings, supplemented from reports to VSNET \citep{VSNET}\footnote{
$\langle$http://www.kusastro.kyoto-u.ac.jp/vsnet/$\rangle$}.
The light curve is shown in Figure \ref{fig:vis}.  This figure corresponds
to figure 1 (CR Boo) of \citet{kat01crboo}.

\begin{figure*}
  \begin{center}
    \FigureFile(160mm,220mm){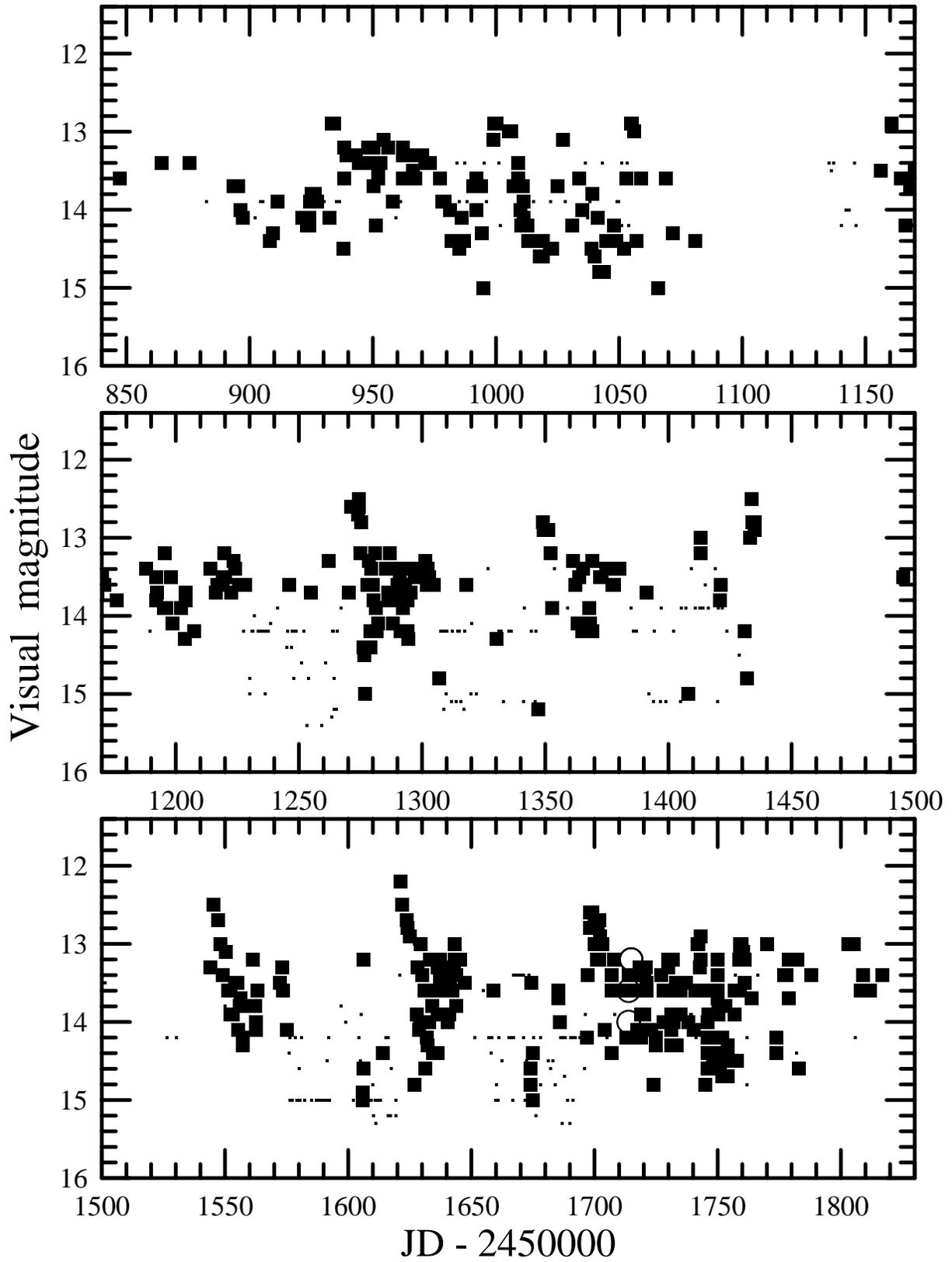}
  \end{center}
  \caption{Long-term light curve of V803 Cen.  The large and small filled
  squares represent positive and negative (upper limit) visual observations.
  The open circles and open squares represent snapshot $V$-band CCD and
  unfiltered CCD (zero point adjusted to $R_{\rm c}$ system) observations.
  Please note that the true quiescence of V803 Cen is below this figure
  ($V\sim$16.8--17.2).  A few CCD observations close to the quiescent
  magnitude were not shown in order to better display the outburst
  behavior.
  }
  \label{fig:vis}
\end{figure*}

\begin{figure*}
\addtocounter{figure}{-1}
  \begin{center}
    \FigureFile(160mm,220mm){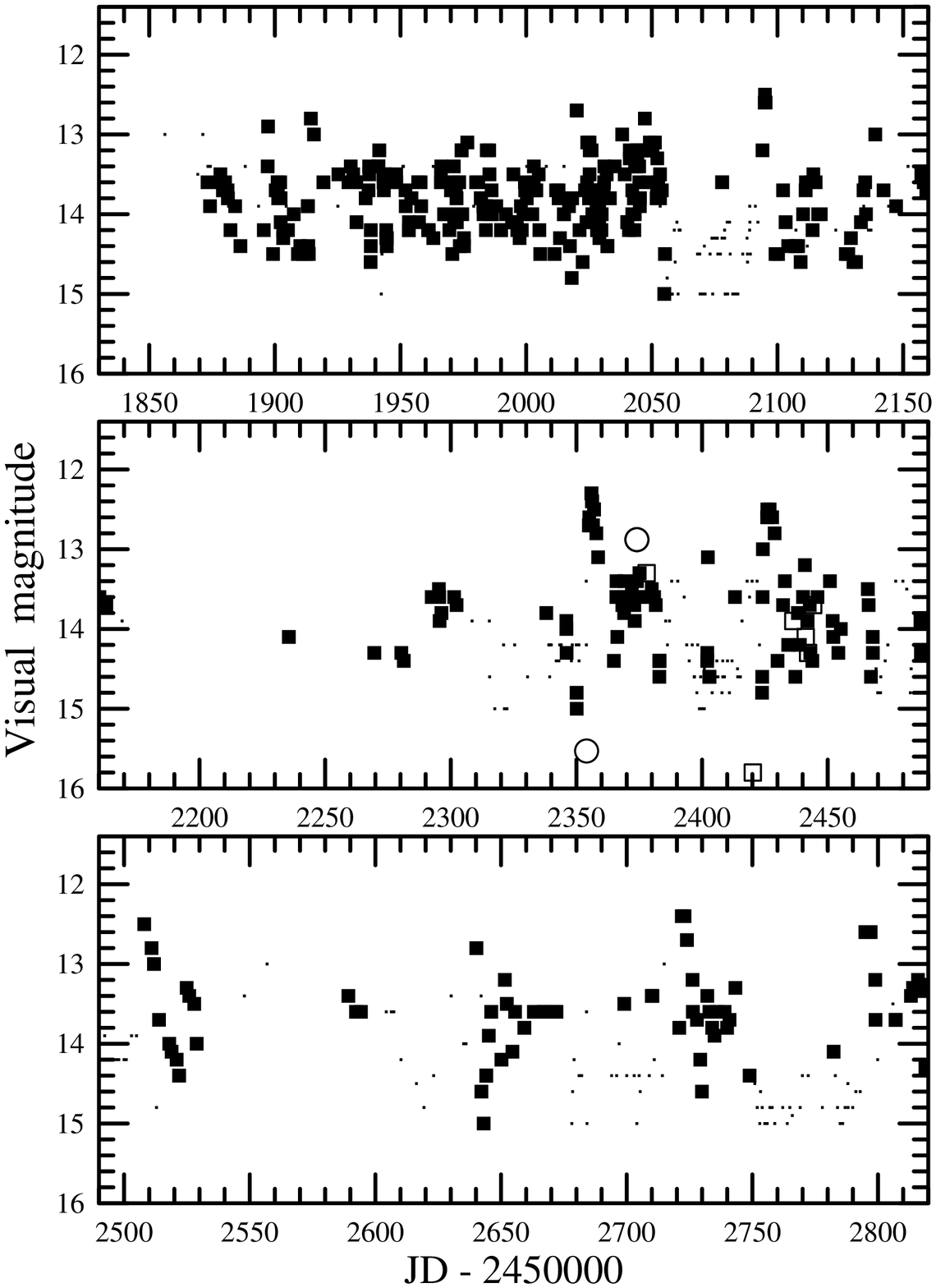}
  \end{center}
  \caption{(continued)}
\end{figure*}

   Upon the detection of the 2003 June bright outburst by Rod Stubbings
(visual magnitude 12.6 on June 4.476 UT, vsnet-campaign-dn 3735,
see also
$\langle$http://www.kusastro.kyoto-u.ac.jp/vsnet/DNe/\\v803cen0306.html$\rangle$),
we initiated CCD photometric campaign through the VSNET Collaboration.

   The CCD observations were performed unfiltered.
Observers and equipment are summarized in Table \ref{tab:equipment}.
Table \ref{tab:log} summarizes the log of observations.
Butterworth and Richards used GSC 7795.1030 for the primary comparison
star, whose constancy during the observation was confirmed by the comparison
with GSC 7795.1402.  Monard and Bolt used the primary comparison star
GSC 7795.184 and the check star GSC 7795.1030.  Systematic corrections
to individual observers were determined both by calculating mean differences
of comparison and check stars, and by maximizing the correlation between
simultaneous observations.  All observations were thus converted
to a common scale (system close to $R_{\rm c}$) relative to GSC 7795.184.
Errors of single measurements were typically less than 0.01--0.03 mag
unless otherwise specified.

   Barycentric corrections to the observed times were applied before the
following analysis.

\begin{table}
\caption{Observers and Equipment.} \label{tab:equipment}
\begin{center}
\begin{tabular}{cccc}
\hline\hline
Observer    & Telescope$^*$ &  CCD  & Software \\
\hline
Butterworth & 20-cm SCT & ST-7E & AIP4Win \\
Monard      & 30-cm SCT & ST-7E & AIP4Win \\
Bolt        & 25-cm SCT & ST-7  & MuniPack \\
Richards    & 18-cm refractor & ST-7E & AIP4Win \\
\hline
 \multicolumn{4}{l}{$^*$ SCT = Schmidt-Cassegrain telescope.} \\
\end{tabular}
\end{center}
\end{table}

\begin{table*}
\caption{Journal of CCD photometry.}\label{tab:log}
\begin{center}
\begin{tabular}{ccrccrc}
\hline\hline
\multicolumn{3}{c}{Date}& Start--End$^*$ & Exp(s) & $N$ & Observer \\
\hline
2003 & June &  6 & 52796.860--52797.072 &  60 & 227 & Butterworth \\
     &      &  7 & 52797.860--52798.079 &  60 & 236 & Butterworth \\
     &      &  7 & 52798.251--52798.473 &  15 & 654 & Monard \\
     &      &  8 & 52798.934--52799.068 &  60 & 149 & Butterworth \\
     &      &  8 & 52798.966--52799.166 &  45 & 336 & Bolt \\
     &      &  8 & 52799.235--52799.492 &  45 & 267 & Monard \\
     &      &  9 & 52800.183--52800.324 &  45 & 162 & Monard \\
     &      & 10 & 52801.057--52801.125 & 180 &  32 & Richards \\
     &      & 10 & 52801.193--52801.508 &  45 & 441 & Monard \\
     &      & 11 & 52802.186--52802.402 &  45 & 312 & Monard \\
     &      & 12 & 52802.972--52803.061 &  60 & 116 & Bolt \\
     &      & 12 & 52803.175--52803.427 &  45 & 308 & Monard \\
     &      & 13 & 52803.981--52804.145 &  60 & 214 & Bolt \\
     &      & 13 & 52804.177--52804.499 &  45 & 456 & Monard \\
     &      & 14 & 52804.938--52805.199 &  60 & 340 & Bolt \\
     &      & 14 & 52805.183--52805.413 &  45 & 326 & Monard \\
     &      & 15 & 52806.230--52806.482 &  45 & 358 & Monard \\
     &      & 16 & 52806.934--52806.985 &  60 &  66 & Richards \\
     &      & 16 & 52807.259--52807.466 &  45 & 289 & Monard \\
\hline
 \multicolumn{7}{l}{$^*$ BJD$-$2400000.} \\
\end{tabular}
\end{center}
\end{table*}

\section{Results}

\subsection{Long-Term Light Curve}\label{sec:long}

   As shown in Figure \ref{fig:vis}, V803 Cen displays at least two
different states of outburst activity.  The intervals JD 2451250--2451700
and JD 2452350--2452820 (possibly since JD 2452090) are characterized by
the presence of interchangeably occurring bright outbursts and a faint phase,
which is sometimes interrupted by short faint outbursts.  This state
is very reminiscent of the 46.3-d supercycle observed in CR Boo
\citep{kat00crboo}.  This state was identified by \citet{kat00v803cen}
to be 77-d supercycle with a superoutburst duty cycle of $\sim$0.4.
The other state covers the interval JD 2451710--2452050 (although there
is an unavoidable seasonal observational gap present, there was no
clear indication of a state change during this interval).  During this
state, V803 Cen was mostly observed between 13.0 and 14.5 mag.
\citet{pat00v803cen} recorded a similar state in 1997 April (outside
this long-term light curve), and referred to this as a ``cycling state"
showing $\sim$1.0 mag variations with a period of 22$\pm$1 hr.
When compared to the 77-d supercycle, this state more looks like
a standstill seen in Z Cam-type (hydrogen-rich)
dwarf novae \citep{kat01v803cen}, although no such cycling variation
was observed during Z Cam-type stars.

   The behavior for the interval JD 2450930--2451070 was rather unusual.
During this period, the first outburst was longer than the typical ones
regularly seen in the 77-d supercycle, and there was an indication of
shortening of outburst intervals to 25--30 d in the later part.  There
is some indication that a similar state was present between JD 2451150
and 2451230.  This state may be something intermediate between the
state with regular 77-d supercycles and the standstill-like state,
as inferred from the duration or duty cycle of the outbursts.  This
state may be analogous to the transient appearance of a 14.7-d period
in CR Boo \citep{kat01crboo}.

\subsection{Overall Light Curve of the 2003 June Outburst}\label{sec:overall}

\begin{figure*}
  \begin{center}
    \FigureFile(160mm,100mm){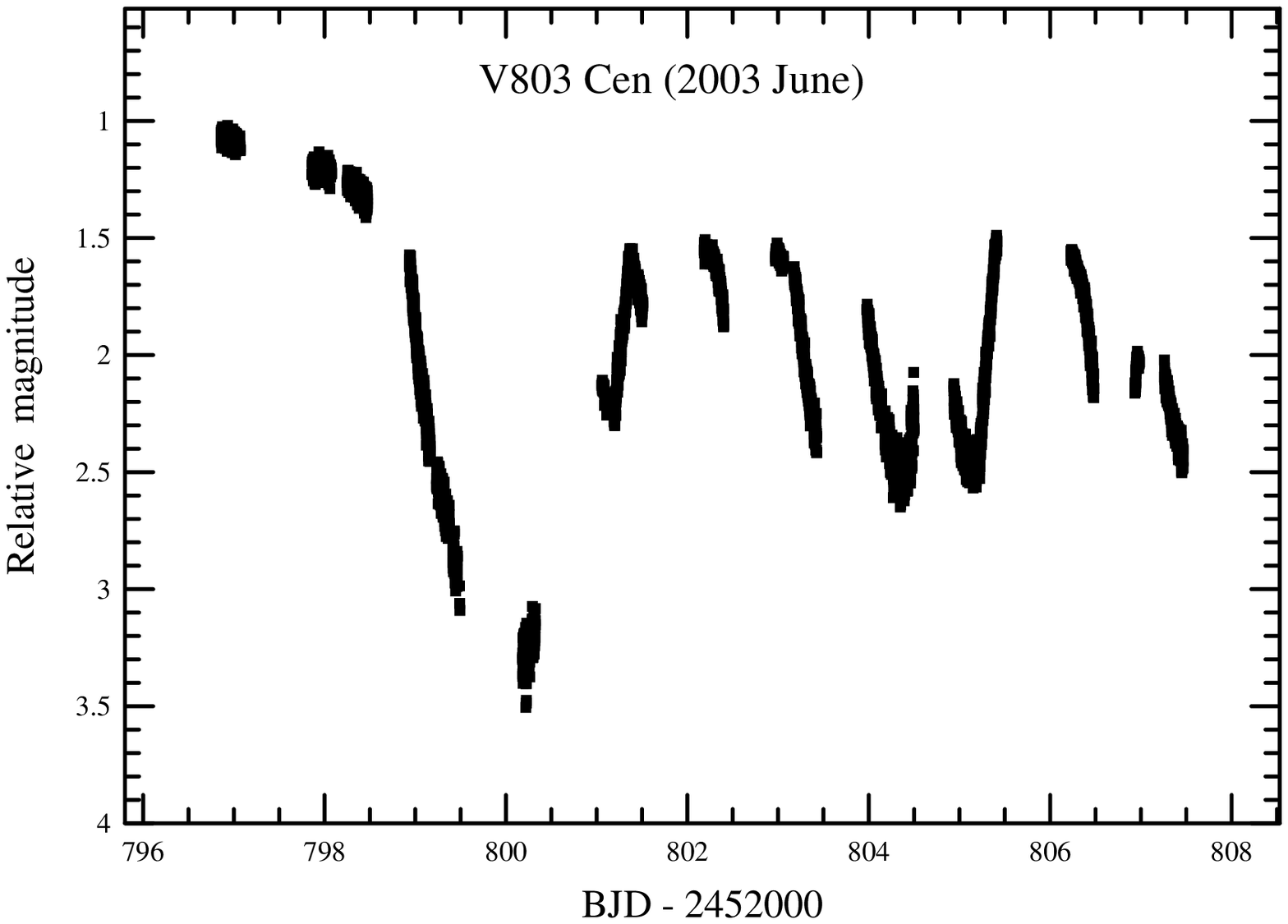}
  \end{center}
  \caption{Unfiltered CCD light curve (system close to $R_{\rm c}$) of
  V803 Cen during the 2003 June outburst.  The magnitudes are given
  relative to GSC 7795.184 (USNO B2.0 mean red magnitude 11.7).}
  \label{fig:lc}
\end{figure*}

   Figure \ref{fig:lc} shows the overall unfiltered
CCD light curve of the 2003
June outburst drawn from the data in Table \ref{tab:log}.  The CCD
observations started 1.9 d after the initial outburst detection.  Visual
observations confirmed that the object remained bright until the initial
CCD observation.  The object remained at bright maximum for $\sim$3 d,
and suddenly started fading (JD 2452799, 2003 June 8).  After reaching
a deep transient minimum (or the ``dip", $\sim$2.5 mag fainter than
the maximum) on JD 2452800 (June 9), the object brightened again and
entered an oscillating state.  This oscillating state is similar to the
``cycling state" described by \citet{pat00v803cen}, but is apparently
different in that the present state was preceded by the sequence of an
initial outburst peak and a subsequent transient ``dip".  During this
state, the mean magnitude gradually faded at a rate of
$\sim$0.1 mag d$^{-1}$.  This sequence of the outburst
activity composed of an initial bright state, a transient fading (dip)
and short-period oscillating state reminds us of the similar sequence
recorded in the 2001 superoutburst of WZ Sge
(\cite{ish02wzsgeletter}; \cite{pat02wzsge}; see also
$\langle$http://www.kusastro.kyoto-u.ac.jp/vsnet/DNe/wzsge01.html$\rangle$)
during its dip--rebrightening
phase \citep{osa03DNoutburst}.  This sequence, composed of initial short
peak brightness, followed by a dip-like fading, and a long-lasting probably
oscillating phase is commonly seen in visual observations
of other superoutbursts\footnote{
  We refer to bright outbursts in the 77-d supercycles as superoutbursts.
} of V803 Cen (Figure \ref{fig:vis}).  This feature was documented as
``damping oscillations" in \citet{kat00v803cen}, although the present
observation far better clarified the detailed behavior at this stage
of a superoutburst.  Please note that this feature is also common
to CR Boo \citep{kat00crboo}.

\subsection{Initial Peak and Superhumps}\label{sec:peak}

   During the first three days of the superoutburst, the object showed
an almost linear, slow decay.  Such a slow variation
was not observed during the subsequent stages.  The mean fading rate
of this stage was 0.16$\pm$0.01 mag d$^{-1}$.  This value is similar
to the mean fading rates of the superoutburst plateau stages of
hydrogen-rich SU UMa-type dwarf novae (cf. \cite{war85suuma};
\cite{pat93vyaqr}; \cite{kat02v359cen}), and is consistent with the
model calculation of a superoutburst in a helium accretion disk
\citep{tsu97amcvn}.

\begin{figure}
  \begin{center}
    \FigureFile(88mm,120mm){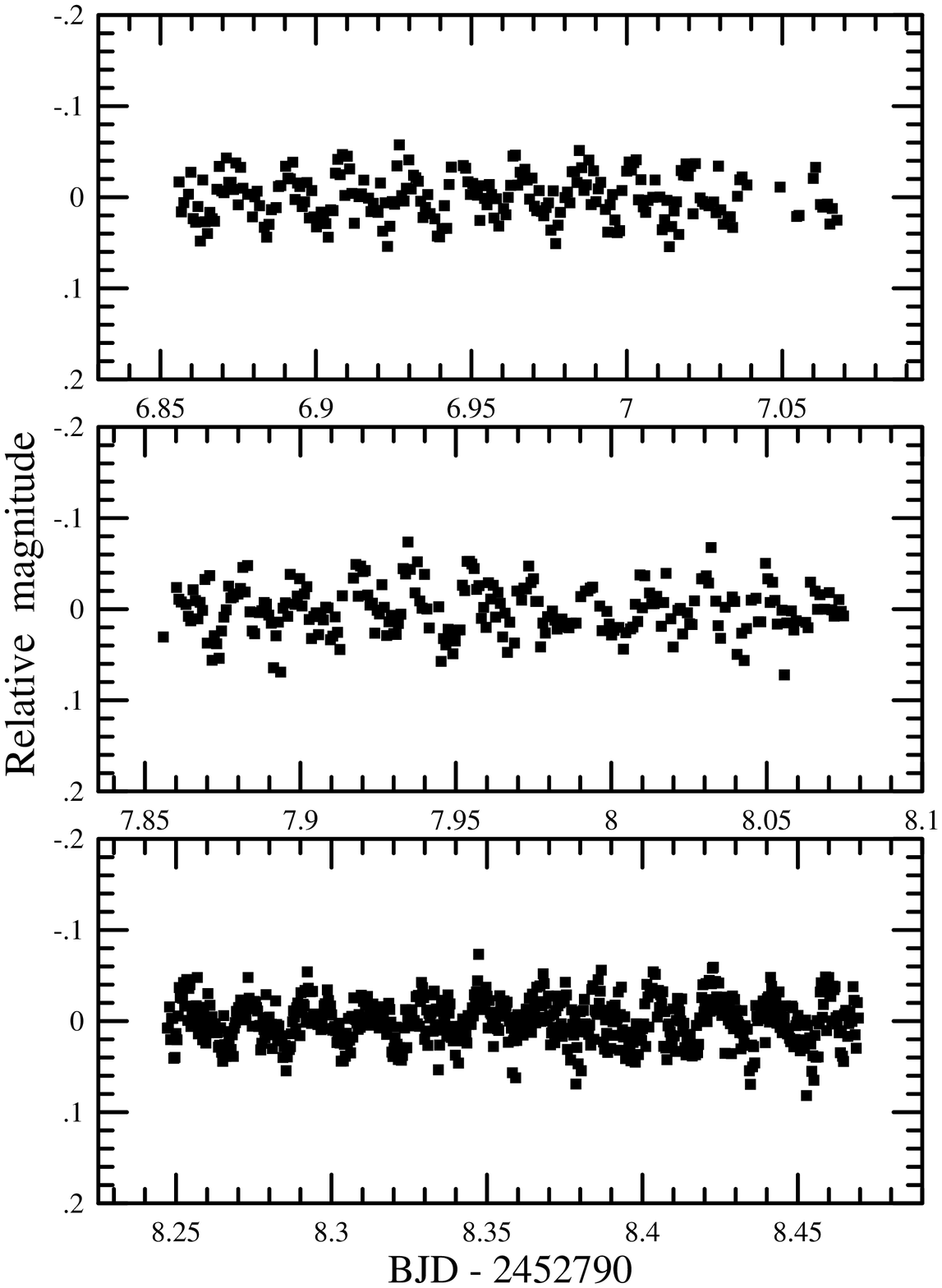}
  \end{center}
  \caption{Superhumps observed during the initial brightness peak of
  the superoutburst.  The slow fading trends were subtracted from the
  observations.  The errors of individual observations are less
  than 0.02 mag.
  }
  \label{fig:peaksh}
\end{figure}

\begin{figure}
  \begin{center}
    \FigureFile(88mm,60mm){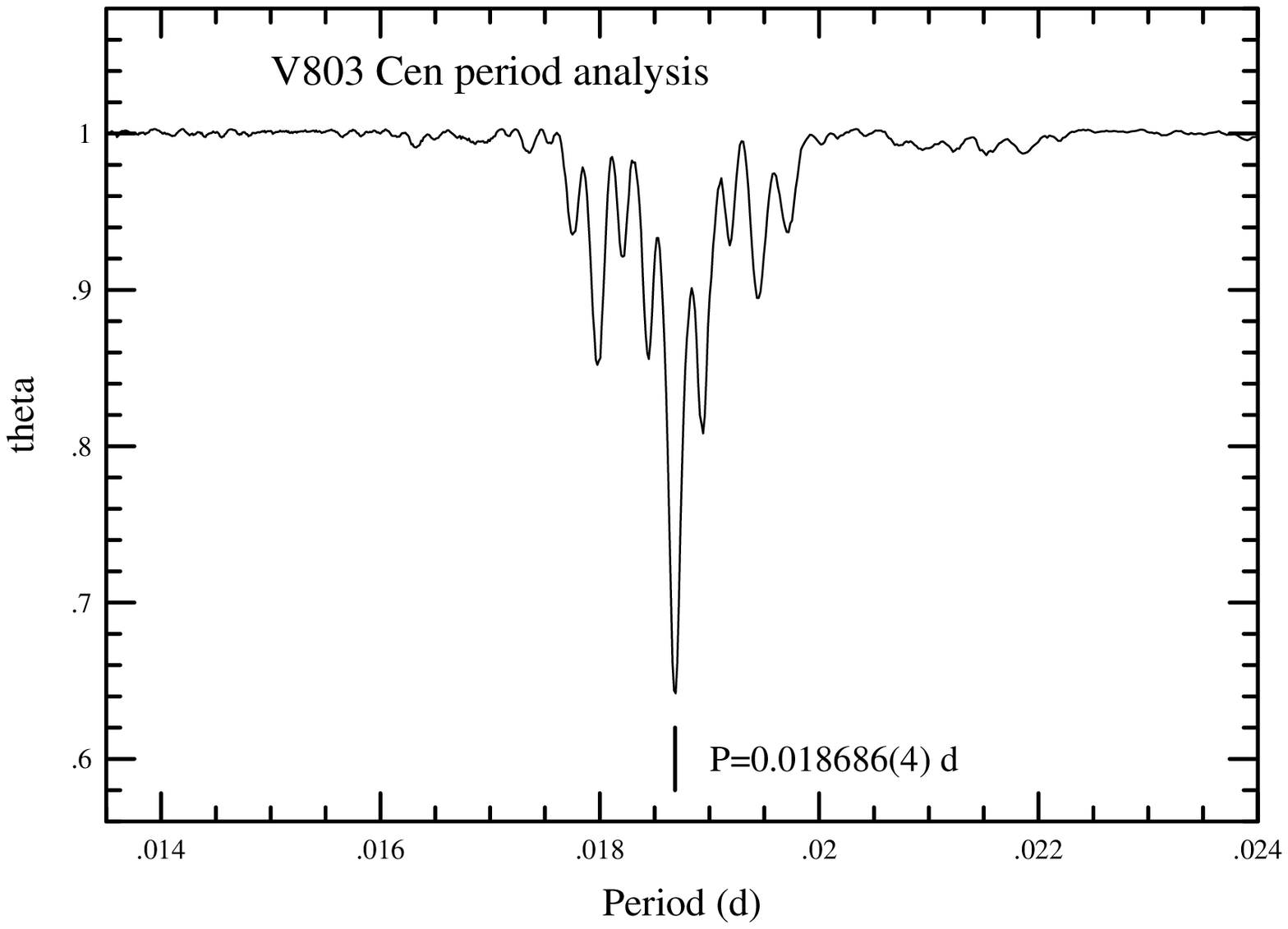}
  \end{center}
  \caption{Period analysis of the superhumps observed during the initial
  brightness peak of the superoutburst.
  }
  \label{fig:pdm}
\end{figure}

   During this stage, clear oscillations were observed
(Figure \ref{fig:peaksh}).  A period analysis of the data during the
initial peak stage with the Phase Dispersion Minimization (PDM: \cite{PDM}),
after removing the linear trend, yielded a period of 0.018686(4) d
= 1614.5(4) s (Figure \ref{fig:pdm}).
The error of the period was estimated using the
Lafler--Kinman class of methods, as applied by \citet{fer89error}.
This period agrees with the 1608--1620 s period reported as the superhump
period (\cite{odo90v803cenWET}; \cite{pat00v803cen}).\footnote{
  \citet{pat02hplib} selected in a table the orbital and superhump periods
  of 1612.0(5) s and 1618.3(8) s, respectively, from various source of the
  literature.  Because of the relatively small fractional superhump excess
  inferred from these periods, and because of the claimed instability of
  the periods \citep{pat00v803cen}, these values should better be treated
  as tentative identifications.
}
The firm presence
of the superhumps as well as the slowly fading phase further confirmed
the superoutburst nature of this initial peak.

\begin{figure}
  \begin{center}
    \FigureFile(88mm,120mm){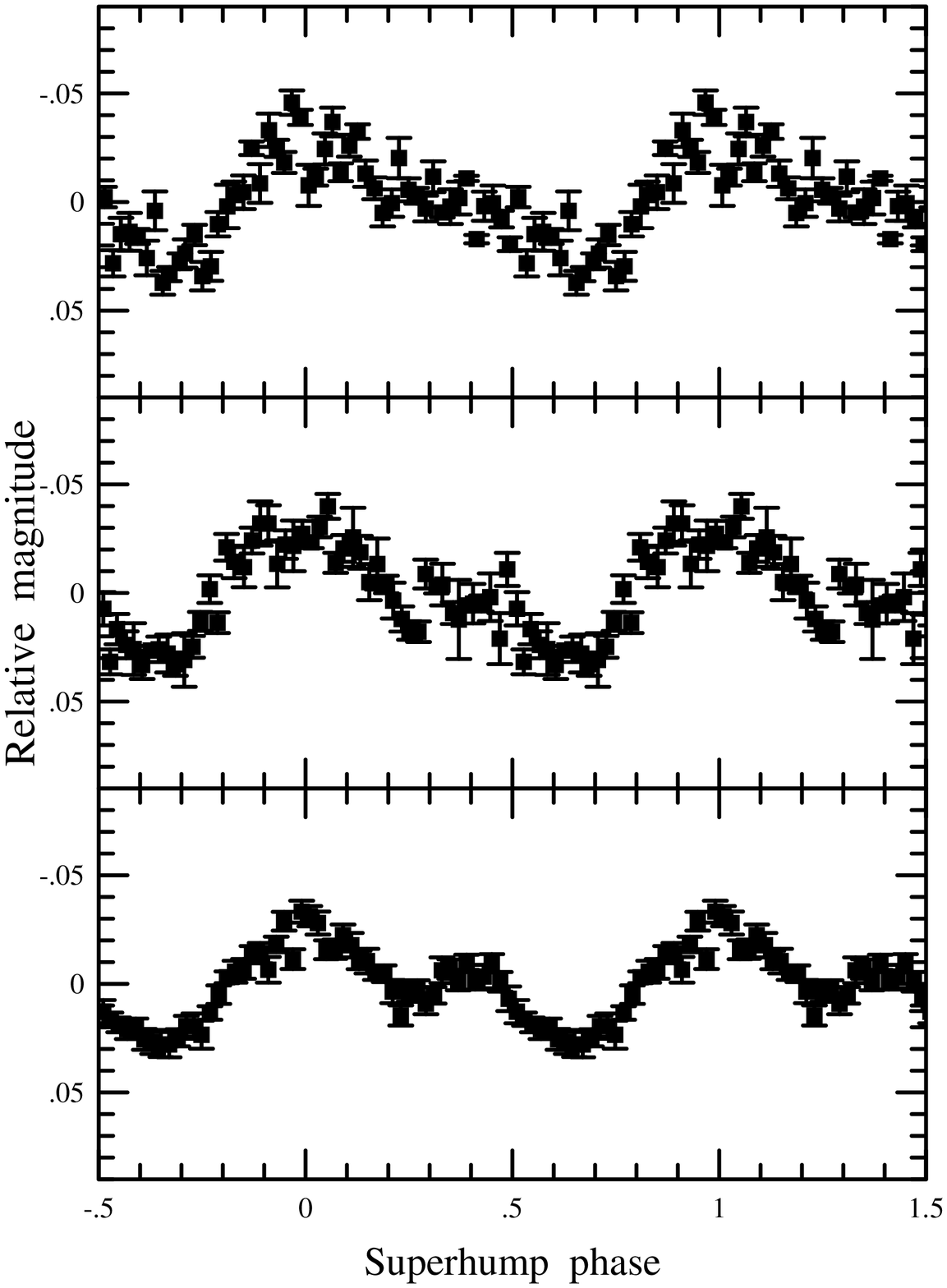}
  \end{center}
  \caption{Mean superhump profiles.  The three panels correspond to the
  three segments displayed in Figure \ref{fig:peaksh}.  The phase zero
  corresponds to BJD 2452789.9950.}
  \label{fig:peakshph}
\end{figure}

   Figure \ref{fig:peakshph} shows the evolution of the superhump profile
during the initial peak phase.  The three panels correspond to the
three segments displayed in Figure \ref{fig:peaksh}.  The superhump
profile was initially triangular, with a rapid rise and a slower decline,
which is characteristic of fully developed (hydrogen-rich)
SU UMa-type superhumps (\cite{vog80suumastars}; \cite{war85suuma}; see
also \cite{odo95amcvn}, \cite{sim98amcvnSH} and \cite{woo95amcvnhplibdisk}
for more discussion of superhumps in IBWDs).
One day later ($\sim$3 d after the initial detection of the outburst),
a secondary maximum appeared in the superhump profile,
similar to secondary maxima is some (hydrogen-rich)
SU UMa-type superhumps (cf. \cite{uda90suuma};
\cite{kat92swumasuperQPO}; \cite{kat03bfara}).  This feature usually
appear at the late stage of a superoutburst, and is possibly related to
``late superhumps" (\cite{hae79lateSH}; \cite{vog83lateSH};
\cite{vanderwoe88lateSH}; \cite{hes92lateSH}), which are known to have
similar periods as ordinary superhumps, but have phases $\sim$0.5 different
from those of ordinary superhumps
The late appearance of
the secondary maximum in V803 Cen follows the ``textbook" evolution of
superhumps of hydrogen-rich SU UMa-type dwarf novae.
The mean superhump profiles displayed in \citet{odo87v803cen};
\cite{odo89v803cen};
\cite{odo90v803cenWET}; \citet{pat00v803cen} more or less showed
this secondary maximum, indicating that the ``pure" triangular superhump
profile is only briefly seen during the early stage of a superoutburst.

\subsection{Rapidly Fading Stage}

\begin{figure}
  \begin{center}
    \FigureFile(88mm,60mm){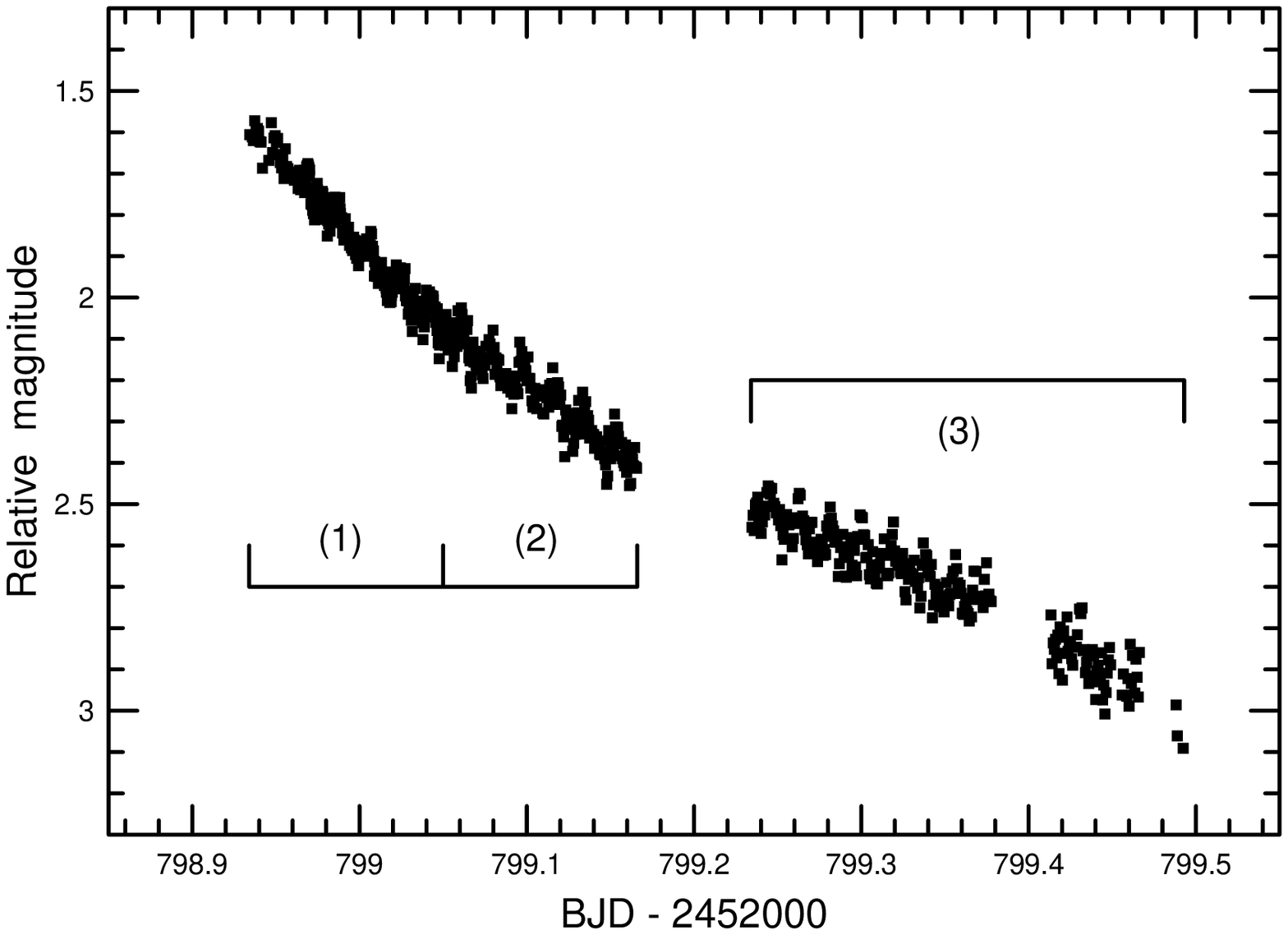}
  \end{center}
  \caption{Enlargement of the rapid decline from the initial peak.
  The segments (1),(2) and (3) correspond to the panels in Figure
  \ref{fig:decph}.
  }
  \label{fig:dec}
\end{figure}

\begin{figure}
  \begin{center}
    \FigureFile(88mm,120mm){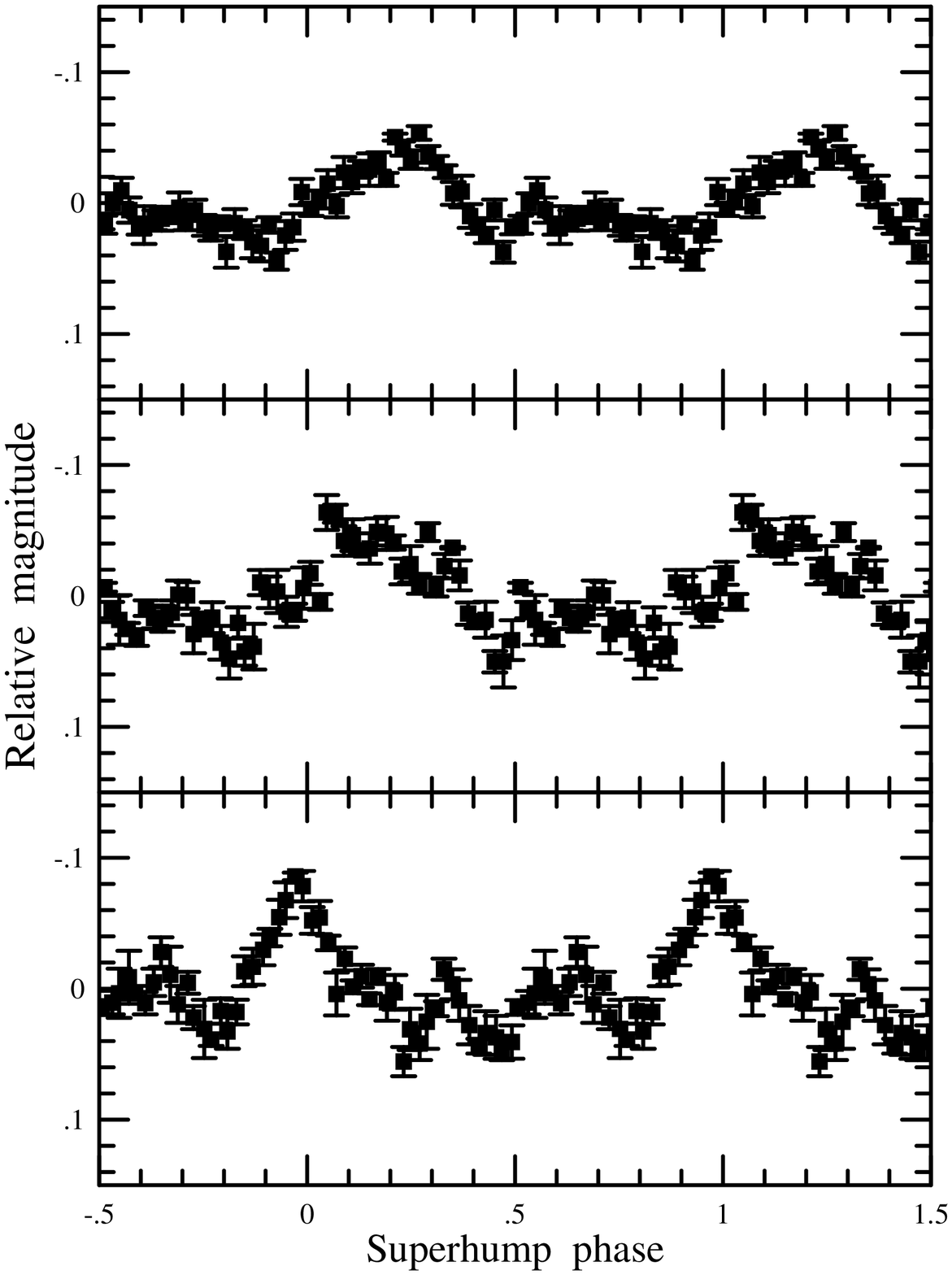}
  \end{center}
  \caption{Mean (super)hump profiles during the rapid decline phase
  assuming the constant period of 0.018686 d.
  The three panels correspond to the segments marked in Figure \ref{fig:dec}.
  The definition of the phase zero is identical with
  Figure \ref{fig:peakshph}.
  }
  \label{fig:decph}
\end{figure}

   Immediately following the initial peak stage, the object started to
fade very rapidly.  Figure \ref{fig:dec} shows the enlargement of the
rapid decline phase.  The mean rates of declines in segments (1), (2)
and (3) are 4.3, 2.6 and 1.9 mag d$^{-1}$, respectively.  Even during
this rapid decline phase, oscillations were clearly present, as is
already evident in Figure \ref{fig:dec}.  Figure \ref{fig:decph} shows
the phase-averaged profiles within the respective segments
in Figure \ref{fig:dec} with the assumption of a constant period
of 0.018686 d (subsection \ref{sec:peak}).

   Assuming the 0.018686-d period, the second maximum of the superhump
(subsection \ref{sec:peak}) was dominant during the earliest stage of
the decline (segment 1).
However the initial peak became stronger as the object faded further
(segments 2 and 3).

\begin{figure}
  \begin{center}
    \FigureFile(88mm,120mm){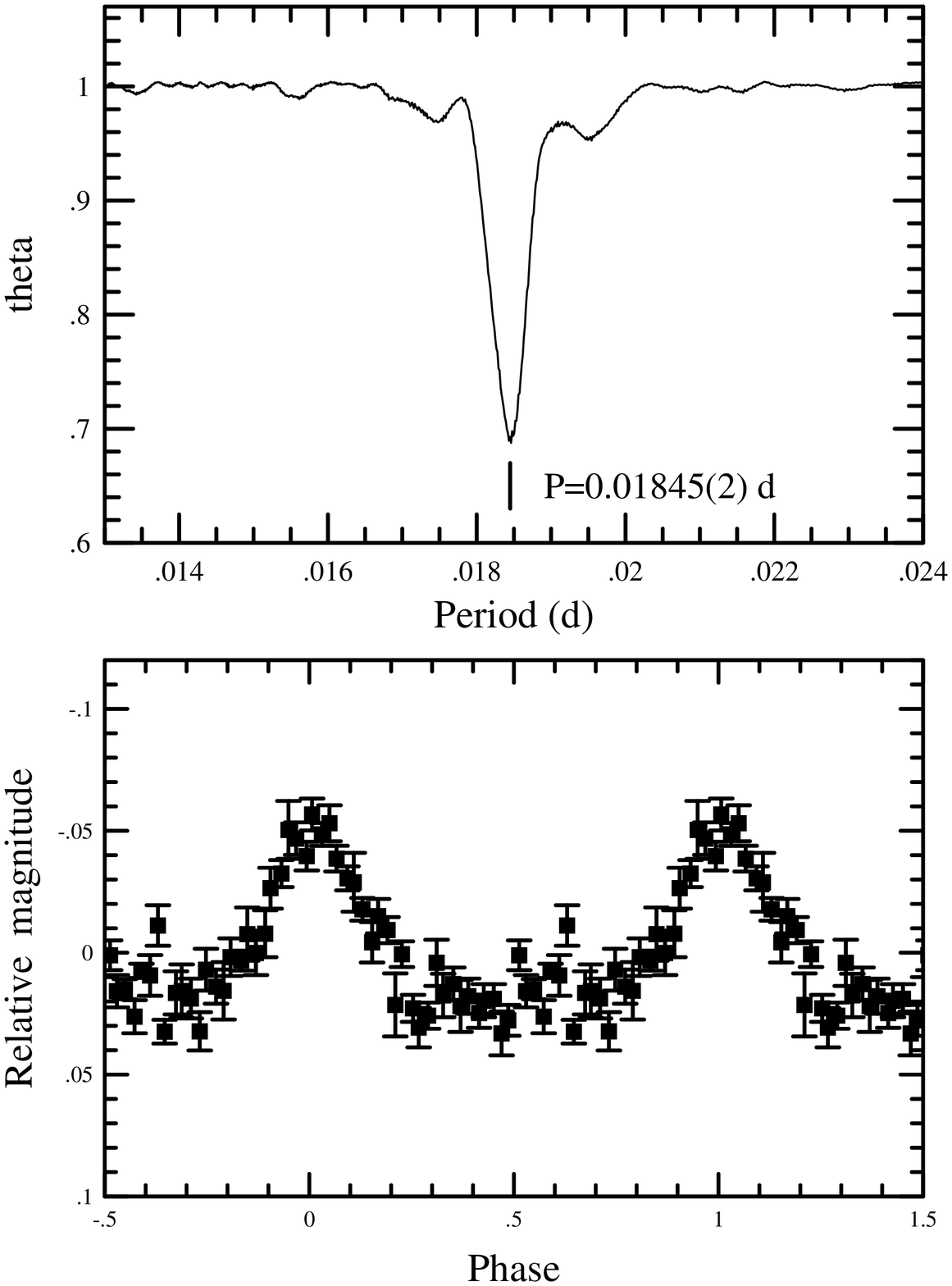}
  \end{center}
  \caption{(Upper:) Period analysis of the rapid decline phase from the
  initial peak.  (Lower:) Mean profile phase-averaged at the best period.
  }
  \label{fig:decpdm}
\end{figure}

   Alternately, we can interpret the progressive variation of the hump
phases as the result of the change in dominant period.  A PDM analysis
of this rapid fading stage (combined set of three segments) has yielded
a period of 0.01845(2) d = 1594(2) s (Figure \ref{fig:decpdm}).
This period is close to the period (cf. \cite{pat00v803cen})
which may be related to the orbital period.  If this period is the orbital
period, the fractional superhump excess becomes 1.3$\pm$0.1\%.
The exact identification of the period, however, should await further
detailed observation fully covering the initial peak through rapid fading
and later stages, and examination of the coherence of the signal over
multiple outbursts.

\subsection{Oscillating Stage}\label{sec:oscillation}

   Following the dip, the object rebrightened into
the oscillating stage, which is remarkably similar to the
{\it rebrightening phase} \citep{osa03DNoutburst} of the 2001 outburst
of WZ Sge (\cite{ish02wzsgeletter}; \cite{pat02wzsge}).
During this stage, the object showed $\sim$1 mag oscillations with periods
of 0.8--1.0 d.  We did not attempt to make an average period of these
oscillations, because the waveforms were apparently variable and a number
of maxima fell within the gaps of observations.

\begin{figure}
  \begin{center}
    \FigureFile(88mm,120mm){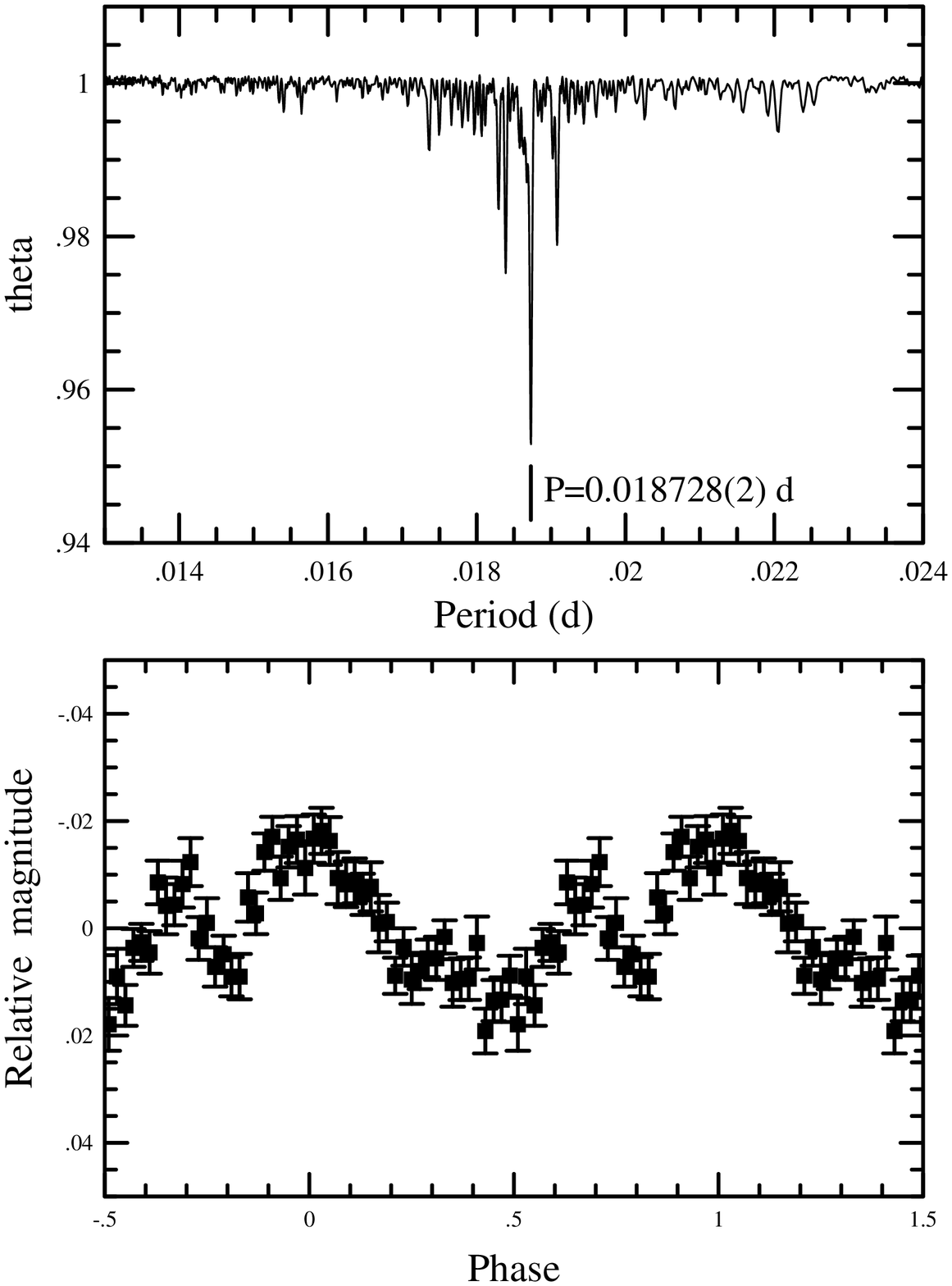}
  \end{center}
  \caption{(Upper:) Period analysis of the short-period variation during
  the oscillating stage.
  (Lower:) Mean profile phase-averaged at the best period.
  }
  \label{fig:rebpdm}
\end{figure}

\begin{figure}
  \begin{center}
    \FigureFile(88mm,120mm){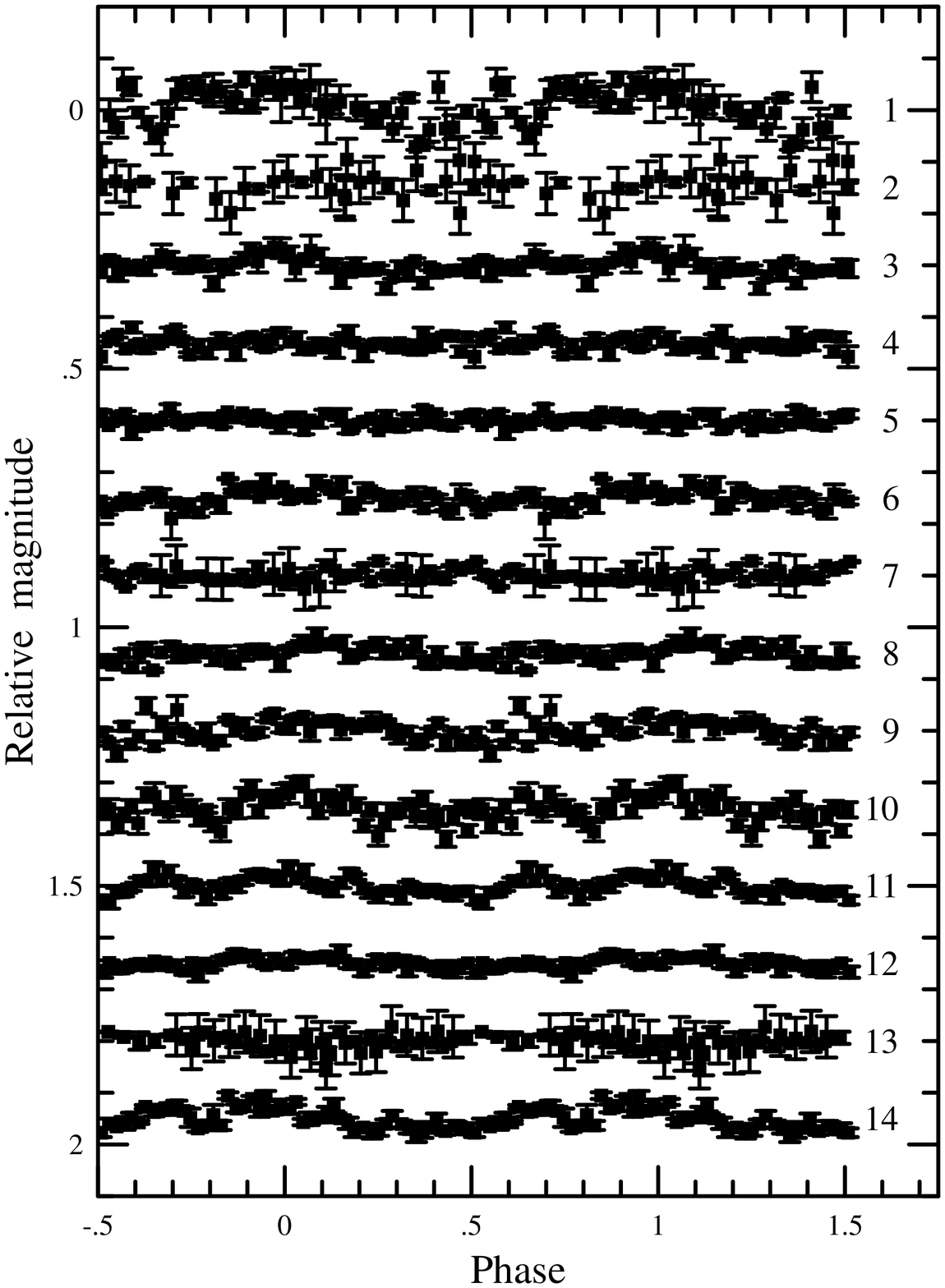}
  \end{center}
  \caption{Profiles of 0.018728-d periodicity during the oscillating
  stage.  The numbers in the right side represent segment numbers
  listed in table \ref{tab:rebseg}.
  }
  \label{fig:phadd}
\end{figure}

   We first prewhitened these oscillations, by subtracting overall
trends with time scales longer than 0.07 d (for slow modulations;
sharp ``kinks" in the light curves were removed by dividing the
light curve into smaller segments) for the interval June 9--17.
A PDM period analysis of the entire oscillating stage yielded a period
of 0.018728(2) d = 1618.1(2) s (Figure \ref{fig:rebpdm}).  The significance
of this period is 92\%.  The mean light curve at this period has a rather
complex structure, with a second maximum around phase 0.7.
Figure \ref{fig:phadd} presents the profile of the 0.018728-d period
(see table \ref{tab:rebseg} for the definition of the segments).
These profiles indicate that this periodicity is variable in amplitude
and in waveform, although no clear evidence of a systematic phase shift
of the peaks was observed.  The signal became weaker when the object
is bright (segments 4--7, 12), and became strongest when the object
is fainter (segments 3, 10, 11, 14).

\begin{table}
\caption{Segments of the oscillating stage used in Figure \ref{fig:phadd}.}
         \label{tab:rebseg}
\begin{center}
\begin{tabular}{cccc}
\hline\hline
Segment & BJD range$^*$ & Mag$^\dagger$ & State \\
\hline
 1 & 0.186--0.324 & 3.3--3.1 & slow rise \\
 2 & 1.057--1.125 & 2.1--2.2 & slow fade \\
 3 & 1.193--1.369 & 2.2--1.6 & rising \\
 4 & 1.370--1.508 & 1.6--1.8 & slow fade \\
 5 & 2.186--2.309 & 1.5--1.6 & near maximum \\
 6 & 2.310--2.379 & 1.6--1.8 & slow fade \\
 7 & 2.972--3.010 & 1.5--1.6 & near maximum \\
 8 & 3.175--3.320 & 1.6--2.1 & fading \\
 9 & 3.981--4.145 & 1.8--2.3 & fading \\
10 & 4.177--4.300 & 2.3--2.5 & slow fade \\
11 & 4.938--5.050 & 2.1--2.4 & slow fade \\
12 & 6.230--6.370 & 1.5--1.7 & slow fade \\
13 & 6.934--6.960 & 2.1--2.0 & slow rise \\
14 & 7.259--7.350 & 2.0--2.3 & fading \\
\hline
 \multicolumn{4}{l}{$^*$ BJD$-$2452800.} \\
 \multicolumn{4}{l}{$^\dagger$ Magnitude relative to GSC 7795.184.} \\
\end{tabular}
\end{center}
\end{table}

\section{Discussion}

\subsection{Overall Outburst Light Curve}\label{sec:outburstdiscuss}

\begin{figure}[t]
  \begin{center}
    \FigureFile(88mm,60mm){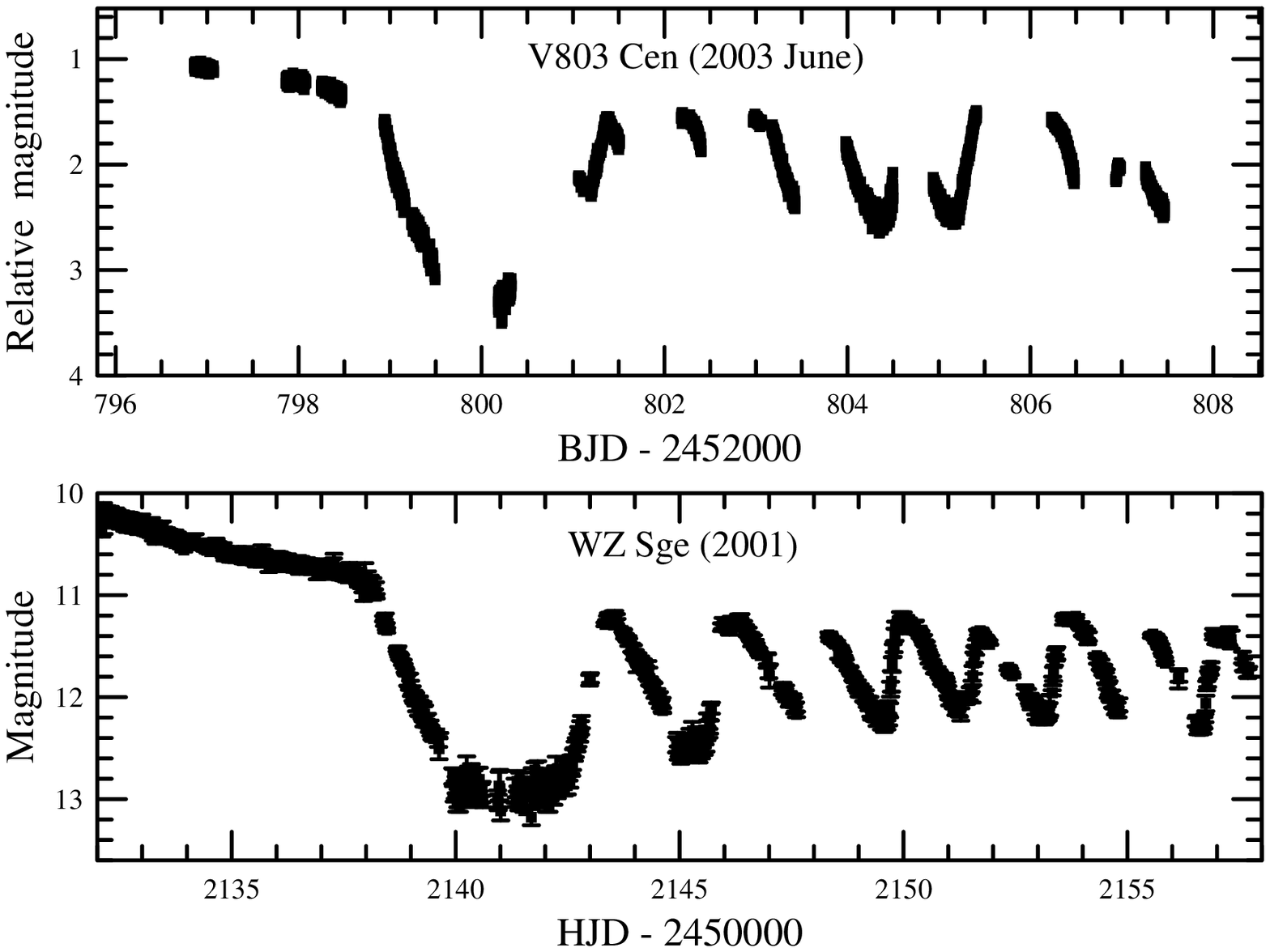}
  \end{center}
  \caption{Comparison of light curves of V803 Cen (2003 June) and
  WZ Sge (2001).  The WZ Sge observations are from \citet{ish02wzsgeletter}
  and R. Ishioka et al. in preparation.  The ``dip" we refer to in WZ Sge
  corresponds to the period HJD 2452139--2452143.
  }
  \label{fig:wz}
\end{figure}

   As mentioned in subsection \ref{sec:overall}, the present outburst,
and possibly past bright outbursts, are composed of a sequence of
an initial bright state, a transient fading (dip) and short-period
oscillating state.  This sequence is very similar to the 2001
superoutburst of WZ Sge (\cite{ish02wzsgeletter}; \cite{pat02wzsge})
during its dip--rebrightening phase (20--45 d after the maximum:
\cite{osa03DNoutburst}).  A comparison of the light curves is given
in Figure \ref{fig:wz}.  The light curve of V803 Cen is repeated here
for clarity.  In V803 Cen, however, the initial superoutburst
phase is much shorter than that of WZ Sge.

   In WZ Sge, this ``rebrightening" phenomenon is interpreted as
a recurring thermal instability (normal outbursts) in the remaining
accretion disk (\cite{osa03DNoutburst}; see also \cite{osa97egcnc},
\cite{osa01egcnc} for the supposed basic mechanism).  In WZ Sge, this
recurrent rebrightening is considered to be maintained by a mass supply
from the outer mass reservoir (see \cite{osa03DNoutburst}; see also
\cite{kat98super} and \cite{hel01eruma} for the basic idea in
SU UMa-type dwarf novae).  As shown in \citet{osa95wzsge} and
\citet{osa03DNoutburst}, a natural explanation of these special features
in WZ Sge-type stars would require an extremely low quiescent viscosity,
which is probably a result of the quenching of the MHD turbulence
under the condition of the low magnetic Reynolds numbers in a cold
quiescent accretion disk of a WZ Sge-type star \citep{gam98}.

   In V803 Cen, however, the large mass reservoir is not naturally
expected as in WZ Sge, since the object has a short ($\sim$77 d)
supercycle, and is expected to have a high mass-transfer
rate ($\dot{M}$) close to thermal stability \citep{tsu97amcvn},
which probably does not satisfy the necessary condition in
\citep{osa95wzsge}.

   The unusual outburst behavior of V803 Cen (and possibly CR Boo),
in turn, may be the result of an extremely small mass ratios
($q=M_2/M_1$, see e.g. \cite{pat02hplib}) and the helium disk.
Considering the extremely short duration [compared to the simulation
by \citet{tsu97amcvn}, who assumed that the superoutburst is
automatically maintained while the tidal torque is above a certain
value] of the initial superoutburst, the superoutbursting state of
the helium accretion disk in V803 Cen may be more difficult to maintain,
thereby resulting in an early quenching of the superoutburst.
This difficulty is to be expected since the helium disk requires
a higher temperature ($T\sim$ 15000 K) to maintain the high state
(\cite{sma83HeCV}; \cite{can84DIdegeneratebinary}).
This early quenching process leaves a substantial amount of
disk matter, which acts similarly as the outer mass reservoir in WZ Sge.

   Alternately, the low tidal torque arising from a small $q$ may be
responsible for the short superoutbursting state.  This explanation
was originally proposed by \citet{nog95rzlmi} to explain the unusual
short superoutbursts in RZ LMi (an extreme ER UMa star,
see \cite{osa95rzlmi}) [see also \citet{hel01eruma} for the application
to other ER UMa stars].  The unusual outburst behavior of V803 Cen
may provide an important link between helium dwarf novae and
unusual hydrogen-rich systems such as ER UMa stars and WZ Sge-type
stars.

\subsection{Standstills and Occasional Shortening of Supercycles}

   As described in subsection \ref{sec:long}, both V803 Cen and CR Boo
show occasional ``standstill"-like states and occasional shortening
of the supercycles \citep{kat01crboo}.  Both phenomena are
difficult to understand in hydrogen-rich SU UMa-type dwarf novae.

   The presence of standstills indicates that the accretion disk is
somehow maintained in the thermally stable state \citep{mey83zcam},
probably by a varying mass-transfer in Z Cam-type dwarf novae.
This mechanism would be, however, less effective in hydrogen-rich
SU UMa-type dwarf novae, since the long-term magnetic activity of the
secondary star, probably responsible for the long-term changes in
$\dot{M}$, is expected to be weaker in a fully convective secondary
of an SU UMa-type dwarf nova.

   There recently exists, however, some observational evidence for
the long-term $\dot{M}$ change even in SU UMa-type dwarf novae
(\cite{kat01v1159ori}; other reported long-term changes of outburst
parameters in SU UMa-type including \cite{kat02suuma}; \cite{kat02v503cyg};
\cite{nog03dmlyr}; \cite{nog03var73dra};
\cite{kat03nsv10934mmscoabnorcal86} may reflect similar activity).
The long-term $\dot{M}$ change may be equally present in helium dwarf
novae.  If this is the case, the typical time scale (1--2 yr) of
the interchanges between outbursting and standstill states is expected
to represent the typical activity time scale in a helium white dwarf,
although the mechanism for such an activity has not been clarified.

   The standstills in V803 Cen and CR Boo are not true
standstills comparable to those of Z Cam-type dwarf novae without
detectable outburst activities.  The oscillation periods (0.8--1.0 d)
during these standstills are too short for usual full-disk outbursts
\citep{tsu97amcvn}, but naturally reflect some sort of thermal
disk-instability.  The presence of thermal instability may also
arise from the difficulty in fully maintaining the hot state in
a helium disk.  The clarification of the exact mechanism, however,
would require future full-disk calculations of a helium accretion
disk.

   The occasional shortening of the supercycles is also difficult to
explain, because high $\dot{M}$ produces longer duration superoutbursts,
resulting in longer duration supercycles in hydrogen-rich systems
\citep{osa95eruma}.  This difficulty, however, may be avoided by
early quenching of the superoutbursting state in a helium disk, or
by an extremely small $q$.

\subsection{Superhump Period}

   The exact identification of the superhump and orbital periods in
helium dwarf novae are often complex, particularly when there are
large-amplitude outbursts.

   During the oscillating state, we obtained a period of 0.018728(2) d
= 1618.1(2) s, which is close to what \citet{pat02hplib} identified
as the superhump period.  However, since a large fraction of the
observation by \citet{pat00v803cen} was performed during the oscillating
state (likely during a ``standstill"), this identification needs to
be treated with caution.  During the initial peak state (corresponding
to the superoutburst plateau in a hydrogen-rich SU UMa-type dwarf nova)
we obtained a slightly different period of 0.018686(4) d = 1614.5(4) s.
Based on the analogy with hydrogen-rich SU UMa-type dwarf novae, the
latter period is better expected to reflect the true superhump period
(\cite{vog80suumastars}; \cite{war85suuma}).

   If the currently observed oscillating stage indeed corresponds to
the ``rebrightening phase" in WZ Sge
(subsection \ref{sec:outburstdiscuss}), the periodic variations
observed during this state may represent late superhumps
(e.g. \cite{vog83lateSH}).  Since the period of late superhumps
(especially long-lasting ones as in WZ Sge-type dwarf novae) is found
to be slightly, and often discontinuously, longer than the main
superhump period (e.g. \cite{pat98egcnc}; \cite{kat03egcnc};
R. Ishioka et al. in preparation),
this slightly longer period can be interpreted
as the period of late superhumps, and the true superhump period is
0.018686(4) d = 1614.5(4) s.  The relation between the signal amplitude
and the system brightness (subsection \ref{sec:oscillation}) is also
identical with the late superhumps in WZ Sge-type dwarf novae
(\cite{kat97egcnc}; R. Ishioka et al. in preparation), which also
supports the interpretation of the longer signal as late
superhumps.\footnote{
  Ideally, it is preferable to demonstrate the presence of
  $\sim$0.5 phase change between the main outburst and the oscillating
  stage.  This was not fulfilled because of the ambiguity of cycle
  counts during the ``dip" phase (even the slight difference between
  the two representative periods, 0.01845 d and 0.018728 d as described
  in the text, already introduces more than 0.5 phase uncertainty
  during the 1-d observational gap in the dip stage),
  and the overall weakness of the
  signal during the oscillating stage (Figure \ref{fig:phadd}).
  Confirmation of the phase jump should await a future continous
  observation.
}

   From the present observation, we have not been able to find
conclusive evidence for the orbital period, although a shorter period
than the superhump period may have been transiently recorded
(subsection \ref{sec:peak}).  Correct identification of the true orbital
and superhump periods should await further long-baseline observations.

\section{Summary}

   We observed long-term behavior of V803 Cen, and confirmed that it
displays at least two different states of outburst activity:
(1) state with outbursts with a supercycle of $\sim$77 d, which is
very reminiscent of the 46.3-d supercycle in CR Boo, and (2)
``standstill"-like state with oscillations with a time-scale of
$\sim$1 d.  These two state interchangeably appear with a time-scale
of 1--2 yr.  During the bright outburst in 2003 June, we conducted
a time-resolved CCD photometry campaign.  This outburst was composed
of distinct states: (1) initial peak lasting for $\sim$3 d with
a slow linear decline, (2) a ``dip"-like transient fading,
(3) oscillating (or rebrightening) stage which resembles
the ``standstill"-like state but with a gradual overall fading.
The overall appearance of the outburst closely resembles that of
the late stage of the 2001 outburst of WZ Sge.
During the initial peak stage, we detected large-amplitude
superhump-type variation with a period of 0.018686(4) d = 1614.5(4) s,
and during the oscillation stage, we detected variations with a period
of 0.018728(2) d = 1618.1(2) s but with a smaller and variable
amplitude.  We consider that the former period better represents the
superhump period of this system, while the latter period may be that of
late superhumps.  The overall picture of the V803 Cen outburst resembles
that of a WZ Sge-type outburst, but apparently with a higher
mass-transfer rate than in WZ Sge-type stars.  We suggest that this
behavior may be either the result of difficulty in maintaining the hot
state in a helium disk or the effect of an extremely low tidal torque
resulting from the extreme mass ratio.

\vskip 3mm

The authors are grateful to observers, especially to A. Pearce
and P. Williams who reported many observations to VSNET.
This work is partly supported by a grant-in-aid (13640239, 15037205)
from the Japanese Ministry of Education, Culture, Sports,
Science and Technology.
The CCD operation of the Bronberg Observatory is partly sponsored by
the Center for Backyard Astrophysics.
The CCD operation by Peter Nelson is on loan from the AAVSO,
funded by the Curry Foundation.
This research has made use of the VizieR catalogue access tool.
We are grateful to R. Ishioka for providing the WZ Sge observations
prior to publication.

\end{document}